\documentclass[12pt]{article}

\usepackage{mathtext}

\usepackage[english]{babel}

\usepackage{cite}
\usepackage{epsf}

\begin{document}

\begin{center}
\Large{Behaviour of muonium in synthetic diamond}
\end{center}

\vskip 5mm

\begin{center}
\large{T.N.~Mamedov$^{1}$, V.D.~Blank$^{2}$, V.N.~Gorelkin$^{3}$,
K.I.~Gritsaj$^{1}$, M.S.~Kuznetsov$^{2}$, S.A.~Nosukhin$^{2}$,
V.G.~Ralchenko$^{4}$, A.V.~Stoykov$^{1,5}$, R.~Scheuermann$^{5}$,
S.A.~Terentiev$^{2}$}
\end{center}

\vskip 4mm

\noindent
$^1$~Joint Institute for Nuclear Research, 141980 Dubna, Moscow reg.,
Russia\\
$^2$~Technological Institute for Superhard and Novel Carbon Materials,
142190 Troitsk, Moscow reg., Russia\\
$^3$~Moscow Institute of Physics and Technology,
141700 Dolgoprudny, Moscow Reg., Russia\\
$^4$~Natural Science Center, Institute of General Physics, RAS,
119991 Moscow, Russia\\
$^5$~Paul Scherrer Institut, CH-5232 Villigen PSI, Switzerland\\

\vskip 4mm

\section{Introduction}

Diamond with its unsurpassed mechanical strength, thermal conductivity,
and radiation hardness is a promising material for radiation detectors, for
electronic and optical-electronic components able to withstand high heat and
radiation loads. Great advances have been made over the last years in
technology of manufacturing synthetic single-crystal diamond and diamond
films~\cite{Adam2003,Tuve2007}. The properties of  synthetic diamond
should be comprehensively studied to find out the scope of its practical
application.

Man-made bulk diamond samples (single crystals and films) contain about
(or more than) $10^{17}$~$\rm cm^{-3}$ dissolved hydrogen atoms. Recently,
surface conductivity was observed in a diamond sample saturated in a hydrogen
atmosphere~\cite{Ley2006}. However, conventional methods do not yield 
sufficient information on the behaviour of hydrogen in diamond.
  
Polarized positive muons $\mu^+$ can be used to imitate and study the
behaviour of an hydrogen atom in matter (see, for example,
\cite{Belousov1,Schenck}).  A positive muon in matter may pick up an
electron and form muonium (${\rm Mu}=\mu^+e^-$).  Since the muon mass is
about 1/9 than that of the proton, muonium can be considered as a light
isotope of hydrogen.  Theoretical calculations~\cite{Belousov, Patterson}
show that in semiconductors with diamond crystal structure the muonium may
occupy tetrahedral and octahedral interstitial sites or is localized in the
middle of the axis between two host atoms (bond-centre site).  It was
experimentally found (see ~\cite{Holz1982,Patterson1988}) that in diamond
the muonium occupies the tetrahedral site (commonly referred to as normal
muonium and designated as Mu or $\rm Mu_T$) and the bond-centre site
(commonly referred to as anomalous muonium and designated as $\rm Mu^*$ or
$\rm Mu_{BC}$).  One more $\mu$SR signal corresponding to the diamagnetic
state of $\mu^+$ was observed in experiments ($\mu^+$-state).

In diamond $\rm Mu_T$ possesses isotropic symmetry with a hyperfine constant
$A_{\rm hf}/h = 3711\pm21$~MHz~\cite{Holz1982} 
($A_{\rm hf}/h=3693\pm83$~MHz~\cite{Spencer1984}), which is slightly smaller
than that for muonium in vacuum $A_0/h = 4463302.765\pm0.053
 $~kHz~\cite{Liu1999}.  At the bond-centre site the hyperfine interaction is
anisotropic with axial symmetry relative to the [111] axes of the crystal. 
The hyperfine interaction constants of $\rm Mu_{BC}$ are $A_{\parallel}/h =
167.98 \pm 0.06$~MHz and $A_{\perp}/h = -392.59 \pm
0.06$~MHz~\cite{Holz1982}.  For the muonium bond-centre site the muon spin
precession frequency in a magnetic field depends on the angle between the
magnetic field and the [111] axes of the crystal.  However, theoretical
calculations~\cite{Belousov,Patterson} reveal that for $\rm Mu_{BC}$ there
is a certain `magic' field where the muon spin precession frequency is
nearly independent on the orientation of the crystal.  This effect can be
used to observe a $\rm Mu_{BC}$ signal at the `magic' field in a powder
sample.

The above-mentioned values of the hyperfine  constants of the muonium were obtained
for two natural diamond samples: Ia-type single-crystal diamond and IIa-type
diamond powder with grain size 1--6~$\rm \mu m$. In the present
work the behaviour of the positive muon in two synthetic diamonds is
studied.

\section{Measurements}

The measurements were carried out at the GPS and Dolly spectrometers
located in the $\pi$M3.2 and $\pi$E1 muon beams of the Paul Scherrer
Institut (PSI, Switzerland).  In the $\pi$M3.2 beam line the muons were
transversely polarized --- the angle between the muon spin and the muon
momentum was $\sim 70^\circ$.

\begin{figure}[htb]
\begin{center}
\fbox{\epsfxsize=6.5cm\epsfbox{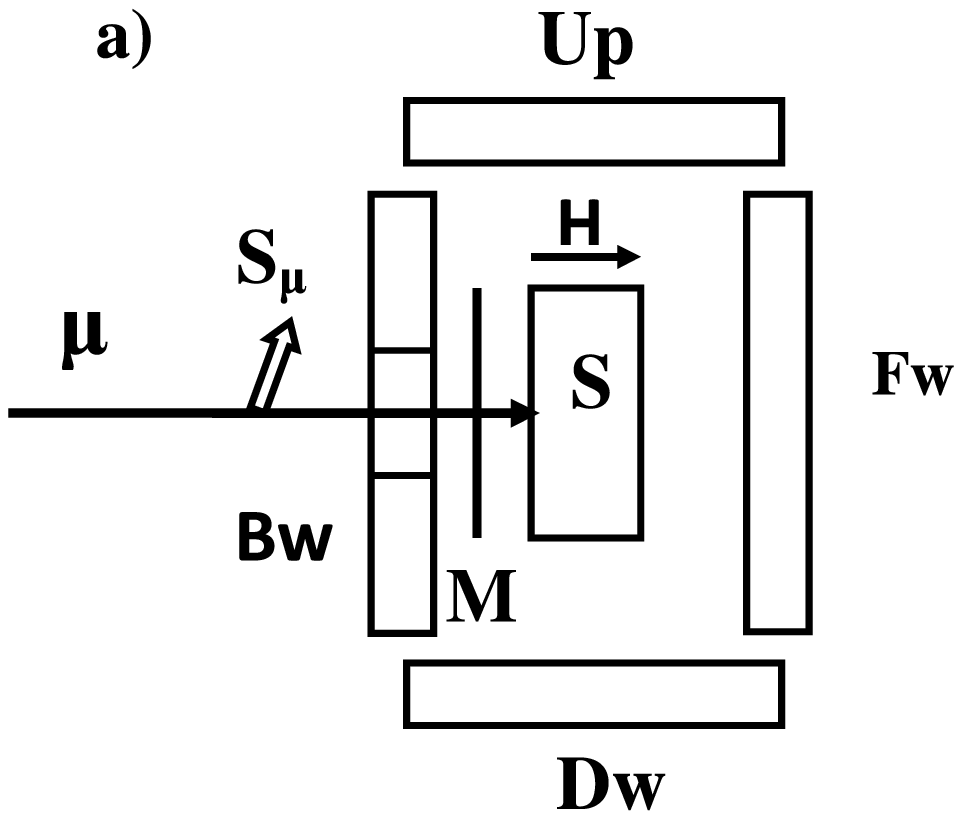}}
\hspace{1cm}
\fbox{\epsfxsize=6.5cm\epsfbox{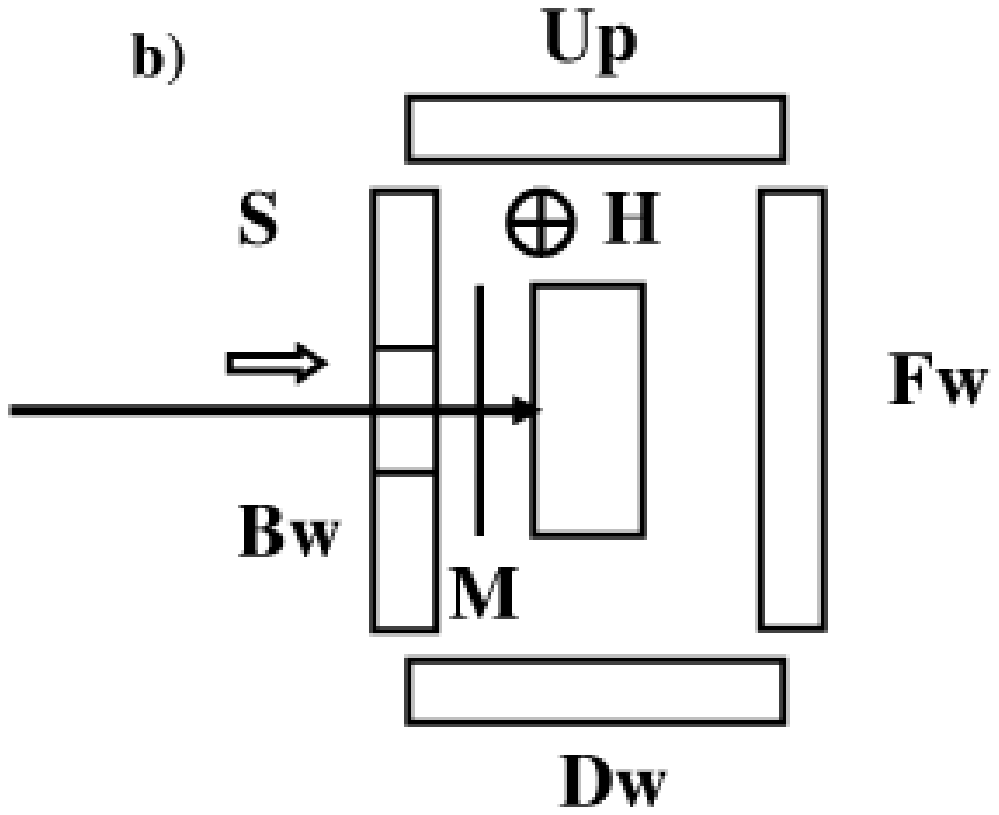}}
\end{center}
\caption{Location of the apparatus in the PSI muon beams.  Up, Dw, Fw and Bw
are the up, down, forward and backward telescopes for detection of
positrons. M is the counter for registration of the incoming muons, $H$ is
the magnetic field, S is the sample, $\rm S_\mu$ indicates the muon spin
polarization direction. a) At the GPS spectrometer the muon beam is
transversely polarized and the magnetic field is parallel to the incoming
muon beam momentum. b) At the Dolly spectrometer the muon beam
longitudinally polarized and the magnetic field is perpendicular to the
picture plane.}
\label{gps-setup}
\end{figure}

The location of the apparatus in the PSI muon beams is schematically shown
in Fig.\ref{gps-setup}.  Muons with energy $\sim 4$~MeV pass through the
hole in the Bw counters and a thin M-counter and stop in the sample S.  The
positrons from the muon decay were detected by the Bw (backward), Fw
(forward), Up(up) and Dw(down) telescopes located around the sample (see
Fig.~\ref{gps-setup}).  A magnetic field transverse to the muon spin
polarization at the sample was created by Helmholtz coils.  The long-time
stability of the magnetic field was better than $10^{-4}$.

Measurements were carried out using a helium flow cryostat that allowed
changing temperature of the sample within the range of 4.2--300~K. The
temperature of the sample was stabilized with an accuracy better than 0.1~K. 
Measurements were carried out in `magic' magnetic field $H_{\rm M}=0.4073$~T at
the GPS spectrometer and in $H=7.5$~mT at the Dolly spectrometer.
 
In these measurements two samples (designated as D3 and D5) were used. 
Sample D3 with the total mass of 2~g was composed of a few IIa-type
single-crystal diamonds. Orientation of the crystals in the sample was
random (MSC -- mosaic single crystal). The concentration of nitrogen in
sample D3 was about 0.5~ppm (in diamond, 1~ppm corresponds to
$1.76 \cdot 10^{17}$~$\rm cm^{-3}$). The crystals contain few wt\% of Ni as
isolated inclusions visible by eye. The diamond crystals with
dimension 1--4~mm were synthesized at the Technological Institute for
Superhard and Novel Carbon Materials (Troitsk, Russia) at high pressure and
high temperature with synthetic diamond as seed material~\cite{Buga2007}.
Sample D5 (CVD) of total mass 2~g was a batch of pieces of several diamond
films 0.5~mm thick and about $10 \times 5$~mm in lateral size.  The diamond
films were produced by a microwave plasma assisted CVD technique in $\rm
CH_4$/$\rm H_2$ mixtures~\cite{Ralchenko97}.  The main impurities in sample
D5 were hydrogen (65~ppm) and nitrogen (1.5~ppm).  Other impurities were
less than 0.1~ppm.  The crystalline axis [110] perpendicular to the film
plane was the predominant grain orientation.

The evolution of the muon polarization $P(t)$ was studied by measuring the
time distribution of positrons from the decay of muons 
$\mu^+ \rightarrow e^+ + \nu_e +\bar\nu_\mu$.
In the general case the time distribution of the positrons
(with respect to muon stop in sample) can be presented by the function
\begin{equation}
N(t)= N_0 \exp(-t/\tau_\mu) (1 + \alpha/3 \cdot G(t)) + {\rm Bg},
\label{Func}
\end{equation}
where $N_0$ is proportional to the number of muons stopped in the sample,
$\tau_\mu$ is the muon lifetime, $G(t)$ is the polarization of the muon at
the decay, $\alpha$ depends on the parameters of the $\mu$SR setup
and is close to 1, $\rm Bg$ is background.

The function $G(t)$ depends on the experimental condition: for
measurements at $H=7.5$~mT and $H=0.4073$~T the respective explicit
expressions for $G(t)$ are
\begin{equation}
\begin{array}{l}
G(t) = P_\mu(0) \exp(-R_\mu t) \cos(2\pi\nu_\mu t+\varphi_0)\\
\qquad {}+P_{\rm T}(0) \exp(-R_{\rm T} t) \cos(2\pi\Omega_1 t+\varphi_1)
\cos(2\pi\Omega_2 t+\varphi_2),
\end{array}
\label{eq:normalMu}
\end{equation}
\begin{equation}
G(t)= P_{\mu}(0) \exp(-R_\mu t) \cos(2\pi\nu_\mu t+\varphi_0) 
+P_{\rm BC}(0) \exp(-R_{\rm BC} t) \cos(2\pi\nu_{\rm M} t+\varphi_1),
\label{eq:anomalMu}
\end{equation}
where $\Omega_1 = (\nu_{23}+\nu_{12})/2$;
$\Omega_2 = (\nu_{23}-\nu_{12})/2$;
$\nu_{ik}$ is the frequency of the transition between levels
of hyperfine structure of $\rm Mu_T$; variables with indices
$\mu$, $\rm T$ and $\rm BC$ refer to the muon in the diamagnetic,
normal muonium and anomalous muonium states respectively; 
$P_i(0)$ is the muon polarization at $t=0$; $R_i$ is
the relaxation rate of the muon spin; $\nu_{\rm M}$ is the muon spin precession
frequency for the $\rm Mu_{BC}$ fraction in the `magic' magnetic field;
$\varphi_i$ is the initial phase of the muon spin precession.

\section{Results and discussion}

\begin{figure}[htb]
\begin{center}
\fbox{\epsfxsize=7.5cm \epsfbox{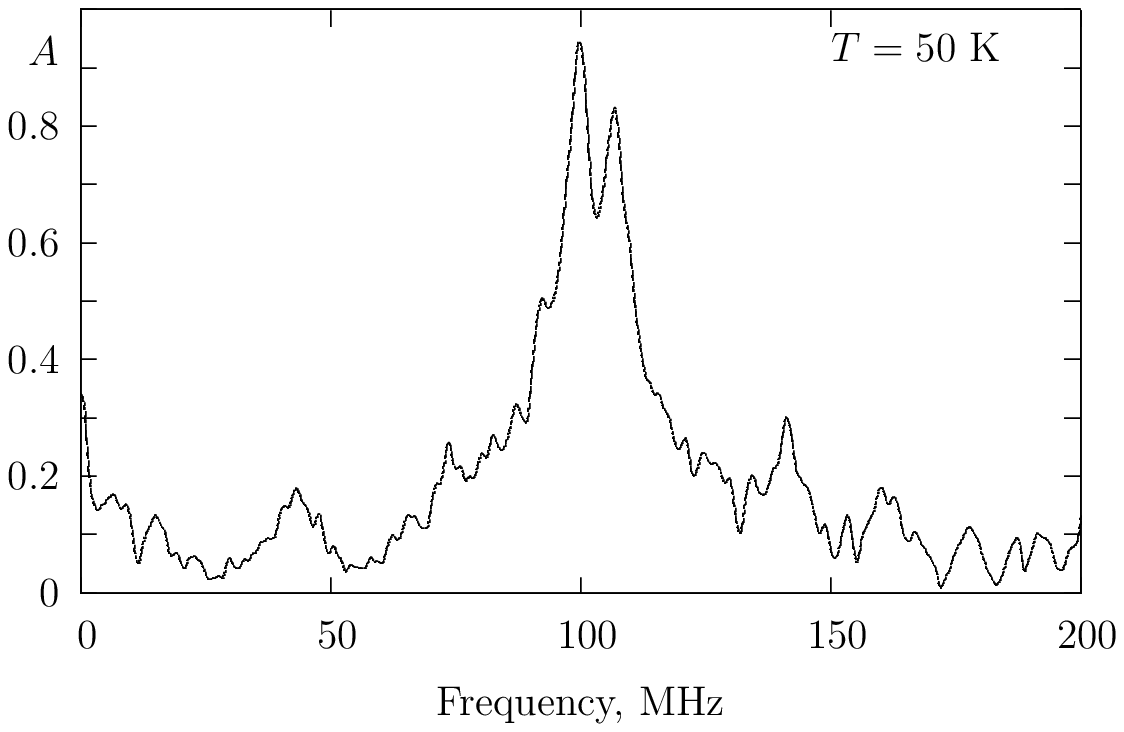}}
\fbox{\epsfxsize=7.5cm \epsfbox{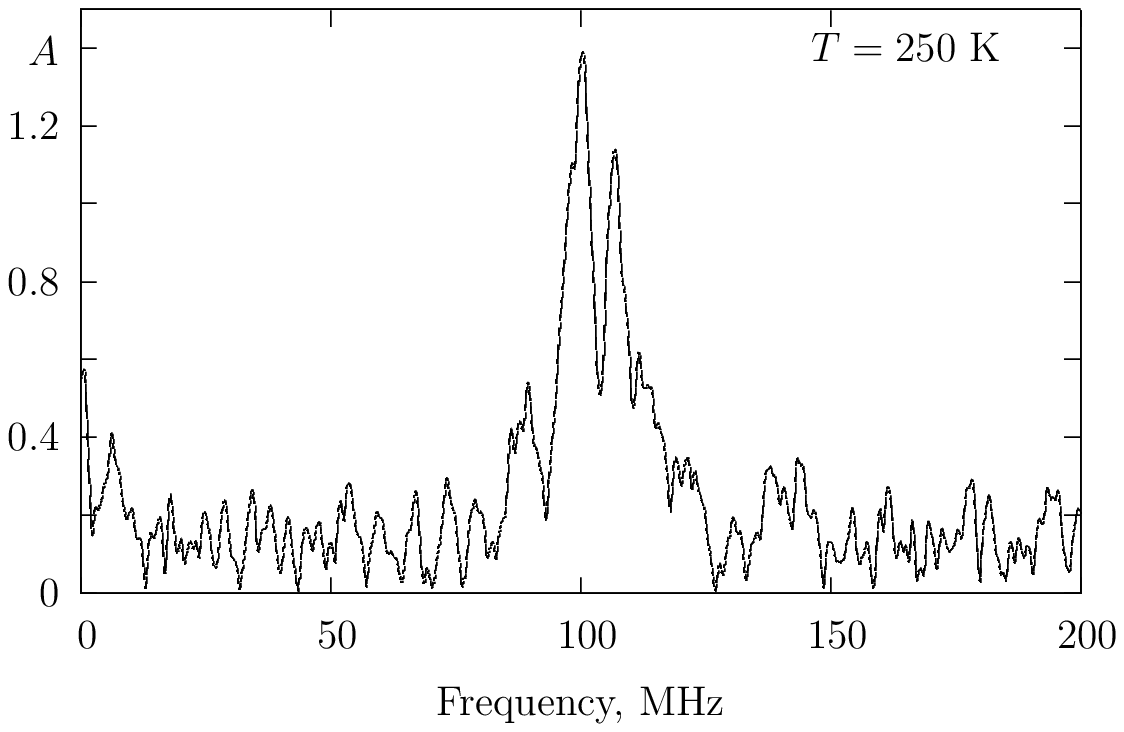}}
\caption{The result of the Fourier analysis of the $\mu$SR  histograms for
diamond film D5 measured in 7.5~mT magnetic field  at the temperatures
a) 50 and b) 250 K.}
\label{fig:CVD_fft}
\end{center}
\end{figure}

Figure~\ref{fig:CVD_fft} shows  the results of Fourier transformation of
the $\mu$SR spectra for sample D5 measured in the magnetic field 7.5~mT
at the temperatures 50 and 250~K. The peaks at $\sim 100.4$ and 
$\sim 105.8$~MHz correspond to the transitions between the levels of the
hyperfine structure of the muonium $\rm Mu_T$. The narrowing of the
lines $\nu_{12}$ and $\nu_{23}$ with increasing temperature was observed for
both samples D3 and D5.

\begin{table}[htb]
\begin{center}
\begin{tabular}{|c|c|c|c|c|c|}
\hline
sample    & 50~K  & 100~K  & 200~K  & 250~K\\
\hline
D3 (MSC)  & \multicolumn{3}{|c|}{$19.1\pm0.8$} & $13.4\pm0.9$\\
\hline
D5 (CVD)  & $12.4\pm0.6$ & $17.3\pm1.4$ & $12.3\pm0.6$ & $7.7\pm0.6$\\
\hline
\end{tabular}
\end{center}
\caption{Muon spin relaxation rate $R_T$ (MHz) for the $\rm Mu_T$
fraction in synthetic diamond at different temperatures.}
\label{tab:relax}
\end{table}

The $\mu$SR spectra measured in the magnetic field 7.5~mT were fitted
by~(\ref{Func}) with polarization function~(\ref{eq:normalMu}) and values of
the parameters $P_\mu(0)$, $P_{\rm T}(0)$, $R_{\rm T}$, $\nu_\mu$, $\Omega_1$
and $\Omega_2$ were found (see Table~\ref{tab:relax} and
Table~\ref{tab:Mu-freq}). It was found that $R_\mu$ is equal to zero, and final
values for other parameters were obtained by fitting the experimental data
with fixed $R_\mu = 0$. The values $\Omega_1$ and $\Omega_2$ presented in 
Table~\ref{tab:Mu-freq} correspond to more precise results obtained at
250~K. Hyperfine constants were calculated as
\begin{equation}
A_{\rm hf}= \frac{1}{2}\left[
{\frac{(\nu_{23}+\nu_{12}+\nu_{\mu})^2}{\nu_{23}-\nu_{12}}-
(\nu_{23}-\nu_{12})}\right] = 
\left[\frac{(\Omega_1 +\nu_\mu/2)^2}{\Omega_2}-\Omega_2\right].
\end{equation}

\begin{table}[htb]
\begin{center}
\begin{tabular}{|c|c|c|c|c|}
\hline
sample   & $\nu_\mu$,~MHz  & $\Omega_1$,~MHz & $\Omega_2$,~MHz & $A_{\rm hf}$,~MHz\\
\hline
D3 (MSC) & $0.996\pm0.005$ & $103.73\pm0.12$ & $2.95\pm0.22$   & $3715\pm277$\\
\hline
D5 (CVD) & $0.992\pm0.004$ & $103.13\pm0.06$ & $2.75\pm0.10$   & $3940\pm143$\\
\hline
\end{tabular}
\end{center}
\caption{Muon spin precession frequency and hyperfine interaction constant
$A_{\rm hf}$ for $\rm Mu_T$.}
\label{tab:Mu-freq}
\end{table}

The values of the hyperfine constant for $\rm Mu_T$ in samples D3 and D5
are smaller than  for muonium in vacuum, and they are in agreement with the
value $A_{\rm hf} = 3711\pm21$~MHz for natural diamond powder~\cite{Holz1982}
within the accuracy of the measurements. However, a significant
difference was observed in the behaviour of  muonium in synthetic diamond in
comparison with  IIa-type natural powdered
diamond~\cite{Holz1982}: 1) the muon spin relaxation rate $R_{\rm T}$ for
samples D3 and D5 is approximately ten times higher than for the natural
sample; 2) in distinction to results~\cite{Holz1982},
$\rm Mu_T$ was observed at temperatures higher than 150~K, and at 250~K the
relaxation rate decreases slightly (see also 
Fig.~\ref{fig:CVD_fft}). At the same time the present results for
the relaxation rate in synthetic samples D3 and D5 are close to those observed
in~\cite{Smallman1996} for IIa- and IIb-type natural diamonds.

\begin{figure}[htb]
\begin{center}
\fbox{\epsfxsize=10cm \epsfbox{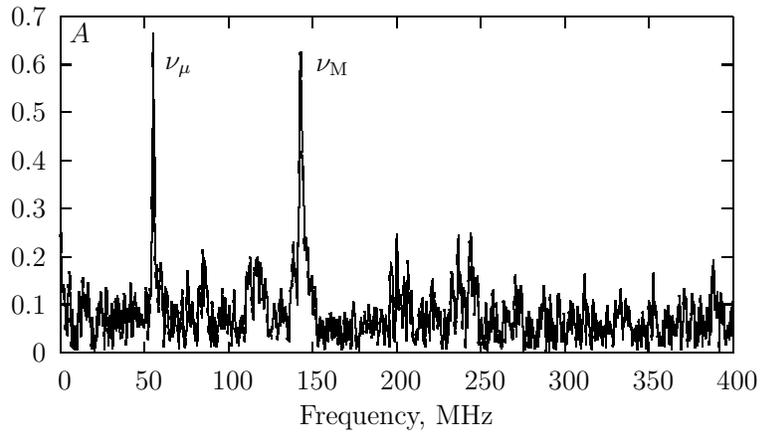}}\\
\caption{Muon spin precession frequencies observed in synthetic diamond D3
in the `magic' magnetic field 0.4073~T at 100~K. }
\label{fig:Mu*fft}
\end{center}
\end{figure}

Figure~\ref{fig:Mu*fft} shows the results of the Fourier analysis of the
$\mu$SR-data measured by the Up telescope of the GPS spectrometer for mosaic
single crystal diamond sample D3 in the `magic' magnetic field 0.4073~T at
100~K.  There are two peaks: one corresponds to the precession of the muon
spin in the diamagnetic state and the other is due to the precession of the
muon spin in the anomalous muonium in the `magic' magnetic field.  Muon spin
precession frequencies of $\rm Mu_T$ too high to be observed in the "magic"
field.  The $\mu$SR histograms collected by the Up and Dw telescopes were
fitted with polarization function~(\ref{eq:anomalMu}) and values of the
parameters $R_{\rm BC}$, $\nu_\mu$ and $\nu_{\rm M}$ were extracted: $R_{\rm
BC}=2.1\pm0.5$~MHz, $\nu_\mu=55.224\pm0.005$~MHz and $\nu_{\rm
M}=142.8\pm0.1$~MHz.  As for the normal muonium in synthetic diamond D3, the
relaxation rate of the muon spin in anomalous muonium is an order of
magnitude higher than that observed earlier in IIa-type natural powder
diamond ($0.25\pm0.10$~\cite{Holz1982}).

\begin{figure}[htb]
\begin{center}
\fbox{\epsfxsize=7.5cm \epsfbox{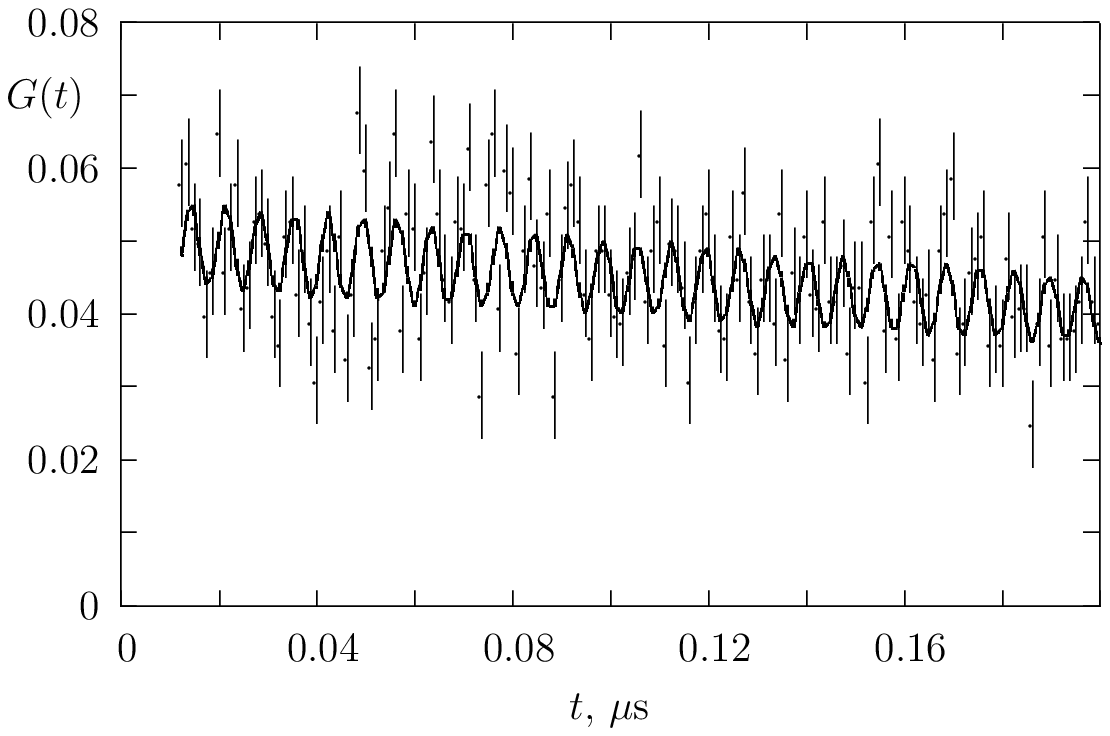}}
\fbox{\epsfxsize=7.5cm \epsfbox{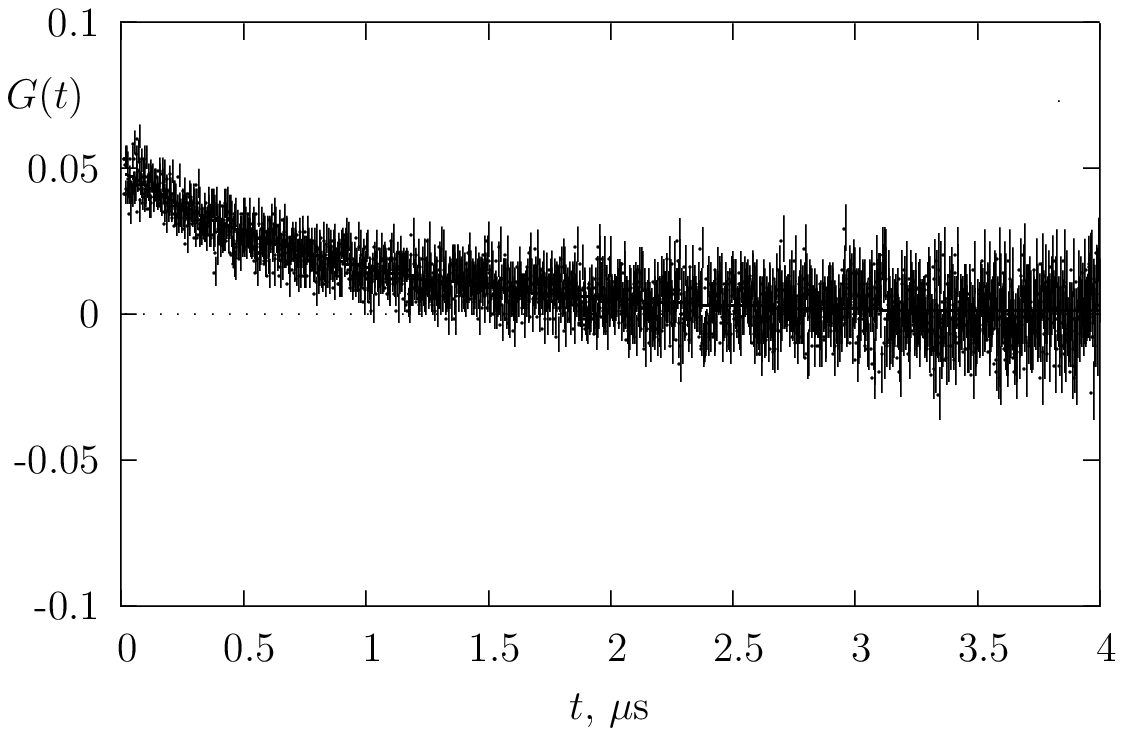}}
\end{center}
\caption{The time dependence of the muon polarization in diamond D3 in the
magnetic field 0.4073~T for the case of longitudinal field measurements
(LF). The data are corrected for the muon life time $\tau_\mu = 2197$~ns.}
\label{fig:longitude}
\end{figure}

Since in the $\pi$M3.2 beam line the angle between the muon spin and the
muon momentum is $\sim70^\circ$ and the external magnetic field is parallel
to the muon momentum, the $\mu$SR-histograms collected by the Fw and Bw
telescopes (see Fig.~\ref{gps-setup}a ) contain information about evolution
of the longitudinal (with respect to the magnetic field) component of muon
polarization (LF measurement).  In Fig.~\ref{fig:longitude} the $\mu$SR
histogram for sample D3 (MSC) collectedted by the Fw telescope at the
magnetic field $H=0.4073$~T and $T=100$~K are presented on two different
time scales.  There are two components of muon polarization in the
histogram: oscillating and non-oscillating.  The Fourier analysis reveals
only one frequency in the Fw and Bw $\mu$SR histograms which corresponds to
muon spin precession of $\rm Mu_{BC}$ in the `magic' magnetic field.
 It is evident that the oscillating component of polarization is due to the
$\rm Mu_{BC}$ fraction.  At the same time both $\rm Mu_T$ and $\rm Mu_{BC}$
fractions may contribute to the non-oscillating component of the Fw and Bw
$\mu$SR histograms.  These histograms were fitted by the polarization
function \begin{equation} G(t) = P(0) \exp(-t/T_1) + P_{\rm BC}(0)
\exp(-t/T_1^{\rm BC})) \cos(2\pi\nu_{\rm M} t+\varphi), \end{equation} where
$T_1$ and $T_1^{\rm BC}$ are the longitudinal relaxation times for the
non-oscillating and oscillating components of muon spin polarization.

The following values of the muon spin polarization parameters 
were found: $1/T_1=1.07\pm0.02$~MHz, $1/T_1^{\rm BC}=1.14\pm0.14$~MHz,
$\nu_{\rm M}=142.68\pm0.06$~MHz. As was expected, the value of the
$\nu_{\rm M}$ obtained from LF measurements is in good agreement with
the results of the TF measurements (data collected by the Up and Dw 
telescopes).
 
To find the fraction of muon formation in the $\mu$, $\rm Mu_T$ and 
$\rm Mu_{BC}$ states in diamond the values obtained for 
$\alpha \cdot P_{\mu}(0)$, $\alpha \cdot P_{\rm T}(0)$ and 
$\alpha \cdot P_{\rm BC}(0)$ were normalized to
the muon polarization for the silver sample measured under the corresponding
experimental condition. It was found that for the Up, Dw and Fw telescopes
of the Dolly spectrometer $\alpha\cdot P_{\rm Ag}(0)/3=0.270\pm0.001$ and 
for the Up and Dw telescopes of the GPS spectrometer 
$\alpha \cdot P_{\rm Ag}(0)/3=0.240\pm0.001$. 
In the calculation it was taken into account that in the transverse magnetic
field only 0.5 of the muon polarization in $\rm Mu_T$ and in `magic'
magnetic field 11/30 of the muon polarization in $\rm Mu_{BC}$~\cite{Holz1982}
gives experimentally observable precession signal. The
fractions of muons formed in the $\mu^+$, $\rm Mu_T$ and $\rm Mu_{BC}$
states in diamond samples D3 and D5 are presented in Table~\ref{Mu-fraction}
in comparison with the analogous data for various diamond samples studied
earlier.

\begin{table}
\begin{center}
\begin{tabular}{|l|l|c|c|c|c|c|c|}
\hline
\multicolumn{2}{|c|}{sample}  & $T$, K & $\mu^+$(\%) & $\rm Mu_T$ (\%) & 
                                    $\rm Mu_{BC}$ (\%) & MF (\%)  & Ref\\
\hline
powder IIa & n & 4.2--90 & $<10$    & 18.5(0.9)   & 9.9(0.7)  & $>60$ &
                                                       \cite{Holz1982}\\ 
powder IIa & n & 296     & $<5$     & -           & 14.5(1.3) & $>80$ &
                                                       \cite{Holz1982}\\
SC Ia      & n & 4.2     & $<10$    & 20(4)       & 11.9(0.9) & $>60$ & 
                                                       \cite{Holz1982}\\
SC    Ia   & n & 5--270  & 4(1)     & 0           & -         & -     &
	                                               \cite{Bharuth1997}\\
\hline
SC IIa     & n & 300     & 8.1(3.0) & 68.9(1.0)   & 22.7(0.8) & 0.3(3.3) &  
                                                       \cite{Patterson1988}\\
SC    IIa  & n & 5--300  & 6(1)     & 61(4)       & 26(3)     & 7(8)  &
                                              \cite{Smallman1996,Bharuth1997}\\
SC    IIb  & n & 5--300  & 14(4)    & 53(4)       & 26(3)     & 7(6)$^*$ & 
                                              \cite{Smallman1996,Bharuth1997}\\
\hline
CVD        & s & 10--300 & $\simeq6$ & $\simeq50$ & $\simeq10^*$ &
                                         $\simeq34^*$ & \cite{Bharuth1997}\\
CVD        & s &  4.5    & 19.4      & 50.2       & 6.8       & 23.6    &
                                                           \cite{Ria2004}\\
SC    IIa  & s &   5     &    4      & 54         & 30        & 12      &  
                                                           \cite{Madhuku2010}\\
CVD        & s & 50--250 & 0.8(0.1)  & 60(1)      & -         & -       & **\\
MSC IIa    & s & 100     & 1.5(0.1)  & 57(4)      & 8.1(0.4)  & 33.4(5) & **\\
\hline
\end{tabular}
\end{center}                                                                            
\caption{Fractions of muon states ($\mu^+$, $\rm Mu_T$, $\rm Mu_{BC}$)
and missing fraction (MF) observed in samples of natural (n) and synthetic
(s) diamonds. SC is single crystal diamond, and MSC is a sample composed from
few single-crystal, CVD is chemical vapor deposited diamond film.
*-we estimated according to the data presented in original publication.
**-present data.}
\label{Mu-fraction}
\end{table} 

The data in Table~\ref{Mu-fraction} are divided into the three sets. The
first set includes data for the Ia-type single-crystal and IIa-type powdered
natural diamonds, the second set presents data for the IIa-type and IIb-type
single-crystal natural diamonds and the third set is for the synthetic diamond.  The
$\rm Mu_T$ fraction is minimal and the missing fraction is maximal for the
first set of samples.  For the second set of samples the missing fraction of
muon polarization is close to zero.  In the synthetic samples the $\rm Mu_{BC}$
fraction is 2--3 times smaller than in the IIa- and IIb-type single-crystal
natural diamonds. The large missing fraction in the IIa-type powder sample
(with grain size 1--6~$\rm \mu m$) was explained by diffusion of the 
$\rm Mu_T$ to the crystalline boundaries~\cite{Patterson}.

It was experimentally shown that a large missing fraction of the muon
polarization in the Ia-type diamond is due to one more muonium state in the
nitrogen-rich diamond~\cite{Machi2000,Ria2004}. As is well known, Ia-type
diamonds contain about 0.1\% (1000 ppm) nitrogen impurities and they are
aggregated to large defects forming A- and B-centres. At A-centres two
nitrogen atoms occupied two nearest substitution sites. The calculation
given in~\cite{Machi2000} shows that the muonium may occupy the bond centre
site between the neighboring nitrogen atoms in the case of A-type defects
and the site between the nitrogen molecules in the case of the B-type defects.

The concentration of nitrogen atoms in the IIa- and IIb-type natural
diamond is approximately three orders of magnitude lower than in Ia-type
diamond. The fact that the missing fraction of muon polarization in the IIa-
and IIb-type natural diamond is close to zero indicates that  nitrogen does
not play any noticeable role in formation of  muonium fractions
in these samples.

In the synthetic diamond the concentration of nitrogen is close to that in
the IIb- and IIa-type natural diamond. Therefore it is reasonable to
compare the results for the samples studied in this work with those for IIa-
and IIb-type natural diamond. As is seen from Table~\ref{Mu-fraction}, in
the synthetic samples the $\rm Mu_T$ fraction is comparable, the $\rm Mu_{BC}$
fraction is 2--3 time smaller and the missing fraction is a few times
larger than in the IIa- and IIb-type single-crystal diamonds. The fact that
the missing fraction in the synthetic diamond is 2--3 times smaller than in
the nitrogen-reach Ia-type natural sample and 2--3 times larger than in the
IIa- and IIb-type samples indicates presence of some other defects in the
synthetic diamond in comparison with the natural one. The observed
difference may be due to the high concentration (more than 10 ppm) of
incorporated hydrogen and due to a small crystalline size (a large effective
surface) in the CVD films and due to the presence of Ni includes in sample
D3 (MSC).

\section{Summary}

It is found that the hyperfine constant for interaction of the muon and
the electron magnetic moments on $\rm Mu_T$ and $\rm Mu_{BC}$ in the
synthetic diamond is close to those in the natural diamond. The
longitudinal relaxation rate of the muon spin in the $\rm Mu_{BC}$ state is
measured for the first time: $1/T_1^{\rm BC}=1.07\pm0.02$~MHz. 
The $\rm Mu_T$ fraction of the muon polarization and the relaxation rate 
$R_{\rm T}$ for the synthetic samples are comparable with those for the 
IIa- and IIb-type natural diamond.

The fact that 1)~the missing fraction of the muon polarization in the
synthetic samples is 2--3 times smaller than  in the Ia-type natural sample
and 2)~the $\rm Mu_T$ fraction and the relaxation rate $R_{\rm T}$ for the 
synthetic samples is comparable with those for the IIa- and IIb-type natural
diamond indicates to the low concentration of such defects as A- and
B-centers in synthetic samples. 

We are grateful to the Directorate of the Paul Scherrer Institute for
allowing us to carry out the experiments at PSI.


\end{document}